\documentclass[conference]{IEEEtran}
\IEEEoverridecommandlockouts

\usepackage{cite}
\usepackage{amsmath,amssymb,amsfonts}
\usepackage{algorithmic}
\usepackage{graphicx}
\usepackage{textcomp}
\usepackage{xcolor}

\def\BibTeX{{\rm B\kern-.05em{\sc i\kern-.025em b}\kern-.08em
    T\kern-.1667em\lower.7ex\hbox{E}\kern-.125emX}}
\begin{document}

\title{Federated Learning with Multi-resolution Model Broadcast  
\thanks{The work of E. G. Larsson was supported in part by ELLIIT and the
KAW Foundation.}
}

\author{\IEEEauthorblockN{Henrik Ryd\'en}
\IEEEauthorblockA{{BNEW - Technology \& Strategy} \\
{Ericsson AB}\\
164 83 Stockholm, Sweden \\
\texttt{henrik.a.ryden@ericsson.com}}
\and
\IEEEauthorblockN{Reza Moosavi}
\IEEEauthorblockA{{Global AI Accelerator } \\
{Ericsson AB}\\
164 83 Stockholm, Sweden \\
\texttt{reza.moosavi@ericsson.com}}
\and
\IEEEauthorblockN{Erik G. Larsson}
\IEEEauthorblockA{Dept. of Electrical Engineering (ISY) \\
Link\"oping University \\
581 83 Link\"oping, Sweden \\
\texttt{erik.g.larsson@liu.se}}
}

\maketitle

\begin{abstract}
 In federated learning, a server must periodically broadcast a model  to the agents.
  We propose to use  
  multi-resolution coding and modulation (also known as non-uniform
  modulation) for this purpose. 
  In the simplest instance, broadcast
  transmission is used, whereby all agents are targeted with one and
  the same transmission (typically without any particular favored beam
  direction), which is coded using multi-resolution
  coding/modulation. 
  This enables high-SNR agents, with high path gains to the server, to receive a more accurate model than the low-SNR agents do, without consuming more downlink resources. 
  As one implementation, we use transmission with a non-uniform 8-PSK constellation, where a high-SNR receiver
  (agent) can separate all 8 constellation points (hence receive 3
  bits) whereas a low-SNR receiver can only separate 4 points (hence
  receive 2 bits). By encoding the least significant information in
  the third bit, the high-SNR receivers can obtain the model with
  higher accuracy, while the low-SNR receiver can still obtain the
  model although with reduced accuracy, thereby facilitating at least
  some basic participation of the low-SNR receiver. We show the effectiveness of our proposed scheme via experimentation using federated learning with the MNIST data-set. 
\end{abstract}

\begin{IEEEkeywords}
federated learning, wireless networks, model broadcast, multi-resolution coding, non-uniform modulation
\end{IEEEkeywords}

\section{Introduction}\label{sec:intro}
Increasing concerns for data privacy have motivated the development of decentralized machine learning systems where training data are stored
and processed locally at the edge users (UEs). An
important example is federated
learning (FL), where multiple (possibly large numbers of) agents
participate in the training of a shared global model by exchanging
model updates with a central parameter server (PS).  FL implements an
iterative process where each global iteration, often referred to as
``communication round'', is divided into three phases:
\begin{enumerate}
  \item The PS sends the current model parameter vector $\boldsymbol{\omega}$ (an $N$-dimensional
    vector) to all participating agents. Either the model is sent
    directly, or a differential compared to the previous round is transmitted.
Either unicast digital transmission, broadcast
or multicast transmission can be used when sending the model.

With unicast digital transmission, the server allocates orthogonal
resources to each agent. For each agent a rate and corresponding coding
and modulation scheme are determined. The model, or a differential
update compared to the previously broadcasted model, is quantized and
compressed using a source code. For each agent, a modulation and
coding scheme is applied, tailored to the rate that the channel to
this agent can support. 

With broadcast transmission, the server transmits in such a way that
all agents can decode the model. Since all agents have potentially different
channel path-losses, this can require
the selection of a very low  transmission rate, using heavy error
control coding. An alternative is to use a code that can be decoded
using only partially received bits. For example, if one uses a channel
code (e.g., LDPC) with a pseudo-random structure of its parity check
matrix, or a rateless code, then the agents with small path-loss (and therefore a high
received SNR) may decode the model broadcast after receiving a
relatively small number of bits. In contrast, agents with a higher
path loss will have to receive more parity bits before they are able to decode. In \cite{vu2022energy}, it was recognized that  high-SNR
agents could not only decode the model faster, but also start their
local training (e.g., stochastic gradient) earlier, hence
enabling a reduction in clock frequency of the computation, which in
turn results in lower power consumption.

Multicasting is also possible. For example, in systems that use TDD
massive MIMO technology one can assign the devices to different groups
and give each group a shared uplink pilot \cite{sadeghi2017max}. This effectively results in the
transmission of multiple beams, each beam adapted to serve a
particular group of agents. One can think of such multicasting as a combination of
unicast and broadcast transmission. There can be circumstances where
such grouping is preferable for performance reasons.

\item The agents perform one or more steps of stochastic gradient
  descent on their own (private) training data and obtain a model
  update. Specifically, the $k$th agent computes a gradient update $\boldsymbol{\delta}_k$
  ($N$-dimensional vector).
  
  \item The $N$-dimensional model updates
    $\boldsymbol{\delta}_{1},\ldots, \boldsymbol{\delta}_{K}$ from the
    agents (henceforth $K$ is the number of agents) are sent back to the
    PS. The PS aggregates these updates by adding them together to
    obtain the updated model $\boldsymbol{\omega}$. A typical aggregation
    rule is to compute a weighted sum of the updates, that is, compute the N-dimensional
    total update
    $\sum_{k=1}^{K}{\alpha_k\boldsymbol{\delta}_k} $ for some weights
 $       {\alpha_k}$. 
Nonlinear aggregation rules are also possible, for example to safeguard against malicious  agents \cite{blanchard2017machine}.

\end{enumerate}

The above procedure is repeated until a termination criterion is
met. The termination can for example be when a desirable
training accuracy is achieved, and/or when a pre-determined iteration
rounds have been reached and/or when a global loss function meets the
minimum requirements. The loss function depends on the computational
problem intended to be solved using the FL system, and typically it  represents a prediction error on a test dataset.

\subsection{Contributions} 

In this paper, we are concerned with Step 1 described above, i.e., the
transmission of the model vector from the parameter server to the agents. We propose to use multi-resolution coding and modulation (also known as non-uniform modulation) for this purpose; see Section \ref{sec:Prop Appr} for more details.   We show the effectiveness of our proposal via experimentation using federated learning with the MNIST data-set. 
Our proposed  approach is, to our knowledge, the first of its kind.
In fact, all related work in wireless federated learning
(Section~\ref{sec:rel}) deals with the \emph{uplink} (from agents to the server) as opposed to the \emph{downlink}
considered here.

\subsection{Related Work}\label{sec:rel}

To our knowledge, there is no previous work on efficient  model broadcasting on {downlink}, from the server to the
agents.

In what follows, we summarize   related work that  deals with the {uplink}, from the agents to the server.
Typically, with digital transmission, the agents are
allocated orthogonal resources for their transmissions. 
Before transmission they may apply gradient compression, sparsification, error-correcting coding,
and digital modulation. Optimized schemes for this have been studied extensively \cite{Alistarh2017QSGDCS,lin2017deep}.
The number of
resource elements (REs) allocated to a specific agent may be adapted
to the length of its gradient (after compression), and the
signal-to-noise-and-interference ratio of the channel for the
agent \cite{amiri2020federated,vu2022energy}.

Previous work for the uplink has also considered device scheduling and  resource allocation \cite{gafni2021federated, Yang2020scheduling,yin2023joint}. To aid the scheduling,   different metrics can be used to quantify the data importance \cite{amiriconvergence,luo2022tackling,yang2019agebased} and the
variability of the data distribution at the devices \cite{taik2021data, shen2022variance}. The characteristics of the wireless link can also be exploited to aid such scheduling \cite{salehi2021federated, importance-aware, Ozfatura2021fastF,chen2022federated,shi2021jointD}.
In \cite{Won2023SlimFL}, the authors propose to use superposition coding for global model aggregation and superposition training for updating local models for FL systems utilizing width-adjustable slimmable neural networks (SNN).

\section{Proposed Approach}\label{sec:Prop Appr}

The main problem in Step 1 (transmission  from the  server to the
agents)  is that some agents may have a very high path-loss, such that it takes too many resources to transmit a full-resolution model to the agent on downlink. Yet, in a
heterogeneous-data scenario these high-path loss  agents may have access to very unique
data, such that they cannot be ignored in the learning process. (As
an example of an extreme case, consider handwritten digit recognition
with 100 agents; if 99 of the devices have access to samples of the
digits ``0''-``8'' but only the 100th agent has access to the digit ``9'',
then this agent must be included in the learning no matter the
cost thereof.)

Another, broader problem is the overhead in terms of downlink resources associated with transmission of  the model parameters
to the agents. The network might only have a
certain allocated downlink resource that it can use for the federated learning service. 
Hence
the network should maximize the model performance given its allocated
resources.  Moreover, there might be delay constraints that  make
it unfeasible to a spend long time  sending a model back and forth to
the devices. This can be due to device mobility (e.g. moving into a
coverage hole), device energy consumption constraints (if the device cannot go
into sleep mode), or certain model training constraints at the
parameter server.

\subsection{Non-uniform modulation}\label{subsec:non-unif}
\begin{figure}
    \centering
    \includegraphics[width=0.5\linewidth,angle=-90]{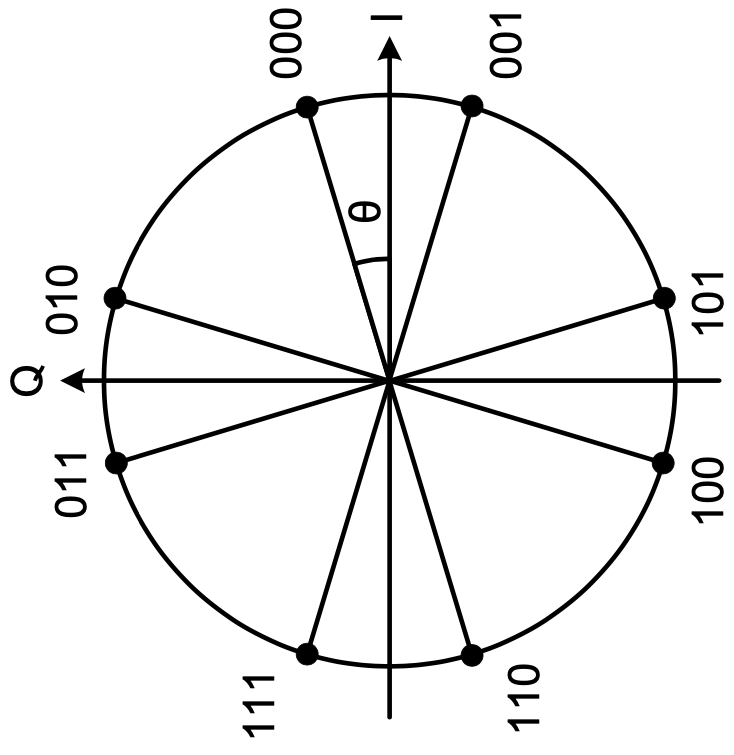}
    \caption{Example of nonuniform constellation for a single-antenna transmitter.}
    \label{fig:nonunimod}
\end{figure}

Various modulation and coding schemes are known by which
high-SNR receivers can decode more bits than low-SNR receivers. The
main intended use case of such modulation and coding in the past has been to facilitate the
transmission of multimedia, on top of voice transmission, when the SNR at
a mobile receiver is sufficiently high. For example, in \cite{pursley1999nonuniform} modulation
with the constellation shown in Figure \ref{fig:nonunimod} is proposed. The constellation
is parameterized through the parameter $\theta$. When received signal is processed at an agent,
the points in this constellation will appear blurred. A high-SNR agent
will be able to separate, with high probability, all 8 constellation
points and therefore detect three bits.  A low-SNR
agent, on the other hand, will with high probability confuse the
constellation points within the pairs (011,010); (111,110); (000,001)
and (100,101). This agent will only be able to reliably decode the first
two of the three bits. 

There are also  non-uniform constellations that can be  used for space-time coded transmission from
multiple-antenna  transmitters \cite{larsson2004nonuniform}. These
constellations are difficult to visualize but they offer the same
basic property: a first stream can be decoded by high-SNR and low-SNR
receivers alike, and a second stream that can only be decoded by
high-SNR receivers.

Returning to Figure~\ref{fig:nonunimod}, the smaller  $\theta$, the larger is the degree of non-uniformity
and therefore high-SNR agents require significantly better channel conditions in order to  detect
the third bit. On the other hand, the smaller the $\theta$, the
smaller is the spread among the received signal points and thus the less is the error probability
for the low-SNR agents. That is, there is a trade-off involved in choosing
$\theta$, between on one hand to offer performance for ``everyone'',
and on the other hand how high the SNR must be to enable transmission
of extra (less significant) bits.
Specific formulas for the error probabilities, quantifying these trade-offs, are available
for single-antenna transmission in \cite{pursley1999nonuniform}.
Corresponding expressions for space-time coded multiple-antenna transmission can be found in
in \cite{larsson2004nonuniform}.

\subsection{Partitioning of the learning model}

In our proposed system architecture, the parameter server obtains a learning model,
represented as a numerical vector $\boldsymbol{\omega}$ of some length
$N$. The components of the vector are normally real-valued. The server
then quantizes $\boldsymbol{\omega}$ into a low-resolution part  $\boldsymbol{\omega}_1=Q\left(\boldsymbol{\omega}\right)$ where
$Q\left(.\right)$ denotes the operation of quantizing to low
resolution. The server furthermore obtains a high-resolution part
  $\boldsymbol{\omega}_2=
\boldsymbol{\omega}-\boldsymbol{\omega}_1$. In practice,
$\boldsymbol{\omega}_2$ may also to be quantized (compressed with loss), and therefore,
$$\boldsymbol{\omega}_1+\boldsymbol{\omega}_2\approx\boldsymbol{\omega}.$$

In principle, any known gradient compression algorithm, here denoted
$Q\left(.\right)$, can be applied to obtain
$\boldsymbol{\omega}_1$. Thereby $\boldsymbol{\omega}_1$ should capture the most
important information (most significant bits) of $\boldsymbol{\omega}$,
while $\boldsymbol{\omega}_2$ captures the least significant part.
In the simplest instance,   the quantization $Q\left(.\right)$
would   operate component-wise, such that
$$\boldsymbol{\omega}_1\left[n\right]+\boldsymbol{\omega}_2\left[n\right]=\boldsymbol{\omega}\left[{n}\right]$$
(where ``$[n]$'' denotes the $n$th component of a vector).

Alternatively, $\boldsymbol{\omega}_1$ could be obtained from
$\boldsymbol{\omega}$ by application of vector quantization. For example,
$\boldsymbol{\omega}$ may be split into several shorter vectors (of
dimension $M$, $M\ll N$) and to each such short vector a vector quantizer
may be applied to obtain corresponding low-resolution parts; the
so-obtained parts are then collated to form $\boldsymbol{\omega}_1$;
finally, $\boldsymbol{\omega}_2$ is obtained.

A specific possibility is to obtain  $\boldsymbol{\omega}_1$   by
retaining the signs of each component of $\boldsymbol{\omega}$, such that
each element is equal to $\pm 1$. Subsequently, $\boldsymbol{\omega}_2$ is then obtained by
computing $\boldsymbol{\omega}_2=\boldsymbol{\omega}-\boldsymbol{\alpha}\boldsymbol{\omega}_1$ for some pre-determined positive constant $\alpha$. This constant in term can advantageously be optimized for maximal 
performance.

\subsection{Partitioning of differential update of learning model}

Typically, the server communicates with the agents during
multiple communication rounds. In each round, the agents provide model
updates, the server aggregates these updates, and transmits the new
model to the agents.

To save on resources, it can be advantageous to transmit only a
differential update of the model. Let $\boldsymbol{\omega}$ be the model
obtained at the server in the current communication round, let
$\boldsymbol{\omega}^\prime$ be the model obtained at the server in the
previous round, let $\boldsymbol{\omega}^{\prime\prime}$ be the model
obtained two rounds ago, and so forth. (Thus $\boldsymbol{\omega}^i$ is the model
obtained $i$ rounds back in time.) Let $\boldsymbol{\omega}^i_1$
and $\boldsymbol{\omega}^i_2$ be the partitioning of the models
into the low-resolution (first) part and high-resolution (second)
part, respectively, for communication round $i$.

Since the low-SNR agents can only decode the low-resolution part of
the transmission, the actual transmitted differential must be based on
the low-resolution part of the model, rather than on the nominal model vector. Hence, at iteration $i$, the
server computes
\begin{itemize}
  \item a low-resolution differential update,
$$\boldsymbol{\omega}_1^i={Q(\boldsymbol{\omega}}^i-\boldsymbol{\omega}_1^{i-1}),$$
    which is decoded by all agents, along with
  \item
    a high-resolution update
$$\boldsymbol{\omega}_2^i=\boldsymbol{\omega}^i-\boldsymbol{\omega}_1^i,$$ which is
    decoded only by the high-SNR agents.
\end{itemize}

Note that differential transmission is only used for the low-resolution part of the
model. This way, the low-SNR agents can keep track of the coarsely
quantized ${\boldsymbol{\omega}_1^i}$ during all communication rounds $i$. The
high-SNR agents can keep track of ${\boldsymbol{\omega}_1^i}$.  
Additionally, since they decode $\boldsymbol{\omega}_2^i$, they can obtain
$$\boldsymbol{\omega}^i=\boldsymbol{\omega}_1^i+\boldsymbol{\omega}_2^i.$$

Before its transmission, the differential update  may first be sparsified such
that a subset of its components are
identified, which are  well
approximated by zero. Consider a general update vector $\boldsymbol{x}$.
Then $\boldsymbol{x'}$ is obtained by
pre-multiplying $\boldsymbol{x}$ with a pre-determined (wide)
compression matrix $\boldsymbol{\Phi}$, such that $\boldsymbol{x}$ can be
recovered from $\boldsymbol{x'}$ with high probability, for example,
by using compressed sensing tools \cite{rish2014sparse}. Subsequently, if such sparsification is applied to
$\boldsymbol{\omega}_1$, 
$\boldsymbol{\omega}_2$ is then
obtained as $\boldsymbol{\omega}_2=\boldsymbol{\omega}- \boldsymbol{\omega}_1$.

\section{Numerical Example}
	In this section, we provide an example of how an FL training process can benefit from having a set of agents being able to receive high-resolution neural network (NN) parameters, while another set of agents can only receive low-resolution NN parameters. The evaluation investigates whether an FL training process can benefit of having a mix of precision for the agents that participate in the same FL process -- for example, enabled by using the multiresolution scheme while broadcasting the model as presented in this paper, instead of adapting the FL model training parameters to the low-SNR agents (by selecting a low resolution for all agents).
 
\subsection{Simulation setting}

The evaluation scenario comprises  4 agents that train a digit classifier model, using the MNIST dataset \cite{lecun1998gradient}. Two of the agents are experiencing high SNR to the base station and can thus both receive $\boldsymbol{\omega}_2$ and $\boldsymbol{\omega}_1$. At the same time, two of the agents   have low SNR and can only receive $\boldsymbol{\omega}_1$. 

The classifier comprises a feed-forward NN according to Table \ref{table_scenario_overview}. The model input is a picture represented by 784 values. Each agent was configured with 2500-digit images each, both for training and testing. The NN uses a 32-bit floating point representation for each NN parameter.  The two high-SNR agents can in our example receive all 32 bits of each parameter, while the two agents with low SNR can only receive with a precision of 2 decimal digits. Note that with the selected 32-bit float, 23 bits are used for the mantissa -- corresponding to roughly 7 decimal digits (the precision for the high-SNR agents).    

As mentioned above, the main point of the example is to show how a training process with FL can benefit by having different precision at each agent. The 2-decimal point precision was selected arbitrarily, and how to optimally select the non-uniform coding that affects the precision at each agent in the non-uniform transmissions remains an open research problem. In general, the selection of such non-uniform coding scheme is dependent on for example the model sensitivity in its trainable parameters, the SNR to each agent, the importance of the training data of each agent, the amount of resources that the NW can spend on model broadcast, and the required model accuracy. It should also be noted that NNs are prominent in adapting to impairments while training, and such impairments can also be used to improve the model robustness. 

\begin{table}[!t]
  \renewcommand{\arraystretch}{1.1}
  \caption{Summary of simulation parameters}
  \label{table_scenario_overview}
  \centering
  \begin{tabular}{|c|c|}
    \hline
    Model parameters &
    Feed-forward neural network with: \\
 & 784-dimensional input,\\
 & 2 hidden layers with \\ & 32 and  64 neurons,  respectively\\
 & ReLU activation layer,\\
 & SoftMax output layer
 \\
     \hline
    Dataset	& MNIST dataset \cite{lecun1998gradient} \\
    \hline
    Number of & \\Training/Test samples	& 2500/2500 for each agent\\
    \hline
    Number of agents	& 4 \\
    \hline
    Model type	& Multi-class classification \\ & with 10 output classes \\
    \hline
    Loss function	& Categorical cross-entropy \\
    \hline
    FL parameters &	1 Training epoch per round,\\& 60 FL iterations,\\& Agents’ parameters are averaged \\& after each iteration.\\
    \hline
    Evaluation scenario & \textbf{1) High resolution} \\
    &All agents receive with 7 decimals precision \\ 
& \textbf{2) Mixed resolution: } \\
&2 agents with 7 decimals precision;\\& 2 agents with 2 decimals precision \\

& \textbf{3) Low resolution:}  \\
&All agents receive with 2 decimals precision
 \\
 \hline
  \end{tabular}
\end{table}

\subsection{Simulation results}

\begin{figure}
    \centering
    \includegraphics[width=1.1\linewidth]{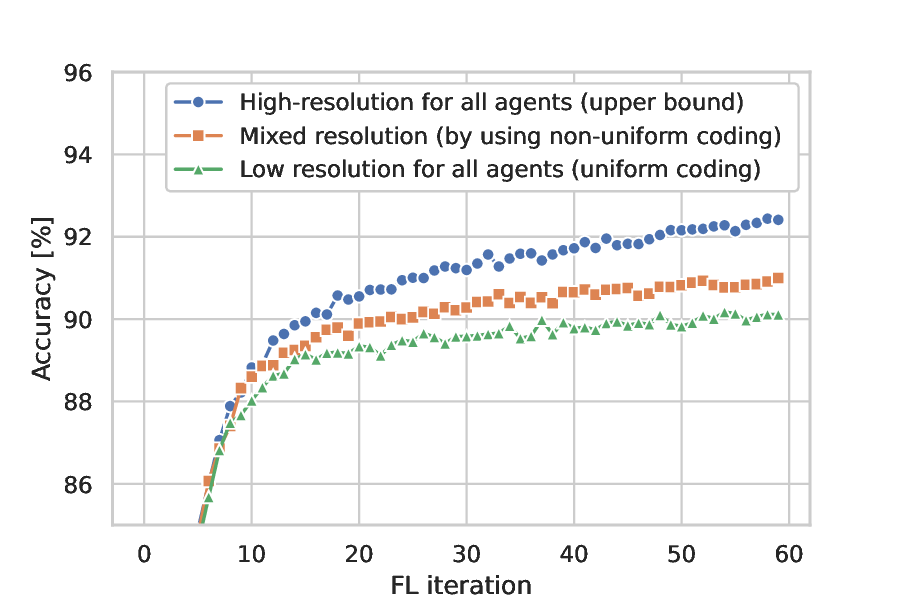}
    \caption{Classification accuracy results for the MNIST test dataset. The model is trained using FL with 4 agents that experience different resolution when receiving the NN parameters.}
    \label{fig:accuracy_plot}
\end{figure}

Due to the dynamic nature of ML model training, for example the impact of random model weight initialization  on the performance, the simulations are carried out with 10 different seeds for each evaluation scenario. The classification accuracy results for the three different schemes are depicted in Figure \ref{fig:accuracy_plot}. The results show that having high resolution for all agents naturally provides the highest performance, roughly 92\% accuracy in classification of  the digits in the test dataset. Using a low resolution for all agents shows the worst performance, around 90\% accuracy. The possible performance while applying non-uniform coding is illustrated in the ``mixed-resolution'' curve, which  shows improved performance in comparison to when all agents receive with the low resolution. The low-resolution scenario is seen as a reference scenario, that corresponds to applying uniform coding, since we need to tailor the precision to the devices with weakest signal quality.

\section{Extensions}

The concepts presented herein can be taken further in a number of different directions.

If some agents have multiple antennas, then they can perform
beamforming to enhance the SNR. This will affect whether they can
decode the high-resolution part or not. Alternatively multiple streams
could be transmitted: a single stream (low-resolution part) to the
single-antenna agents and two streams (low + high-resolution parts) to
the multiple-antenna agents.

Hybrid analog/digital techniques \cite{mittal2002hybrid,shi2011codebook} might be used. For example, the low-resolution
part could be sent using digital modulation (with appropriate error
correction) while the high-resolution part is sent using analog
modulation -- with or without bandwidth expansion.

The non-uniform modulation and coding can as, in the example above,
affect the floating-point value of the model, where less important
bits are transmitted with lower robustness. This could be exploited by transmitting 
each neuron in a neural network with a different probability of
success. Neural networks are in general robust to noise, and the parameter server can decide
to transmit a fraction of nodes with a higher probability of success
than others. This selection could be based on the conductance, as
described in \cite{dhamdhere2018important}.

Extensions to multicast transmission, where the agents are partitioned into groups, could also be developed.  Each group  should then be associated with a set of resources which can include time-, frequency- and beams, and multi-resolution coding should be  applied per group. 

Finally,  in 5G NR,   the
 model updates can be transmitted similarly to system information using
 several synchronization signal blocks (SSBs). One can then directly apply our proposed solution for the
 transmission of model updates. Moreover, there is also a possibility
 of applying different level of multi-resolution coding for different
 transmissions, including sending without multi-resolution coding in
 some of the beams.

\section{Concluding Remarks}

In this paper, we have proposed to use multi-resolution modulation during transmission of a learning model from the parameter server to agents, allowing the agents to receive the model updates with different precision depending on their wireless channel conditions. The proposed concept offers improved resource utilization which in turn translates into improved model performance when having limited network resources available. We showed via an experimentation that the improved model performance resulted in a better image classification.
Another benefit is the increased participation of agents that may have access to, potentially, unique training data even in scenarios with poor wireless propagation (high path loss) on downlink.

The reduced number of resources used for federated learning can reduce the time needed for each learning iteration. For example, the network  use fewer subframes for model multicast, hence enabling a reduction of  the overall time spent on the learning iterations. This could allow agents to go into sleep mode faster, leading to energy savings. Moreover, some agents might not be able to remain during the  entire learning process, due to high mobility (they may move into coverage holes, for example).

\end{document}